\documentclass{ornltm}
\usepackage{threeparttable} % Table with captions and tablenotes
\usepackage{booktabs} % Toprule, midrule, bottomrule
\usepackage[final]{microtype} % Better PDF typesetting
\usepackage{multirow} % Extend table entries over multiple rows

\usepackage{biblatex}  % Numerical style.
\bibliography{mpex.bib} % Replace with your bib filename.
\definecolor{note}{RGB}{48,96,192}

\author{ORNL \and Gary Staebler \and  Rhea Barnett \and Mark Cianciosa \and Rinkle Juneja \and Atul Kumar \and Wouter Tierens \and Minglei Yang \and Cory Hauck \and Richard Archibald \and Pablo Seleson \and Sam Reeve \and LLNL \and Ben Dudson \and Vasily Geyko}

\title{MPEX AI Digital Twins}

\date{\today}

\reportnum{}

\division{Fusion Energy Division}

\begin{document}
\frontmatter
\tableofcontents
%%%%%%%%%%%%%%%%%%%%%%%%%%%%%%%%%%%%%%%%%%%%%%%%%%%%%%%%%%%%%%%%%%%%%%%%%%%%%%%
% ABSTRACT
%%%%%%%%%%%%%%%%%%%%%%%%%%%%%%%%%%%%%%%%%%%%%%%%%%%%%%%%%%%%%%%%%%%%%%%%%%%%%%%
\begin{figure}
  \centering
  \includegraphics[scale=1.0]{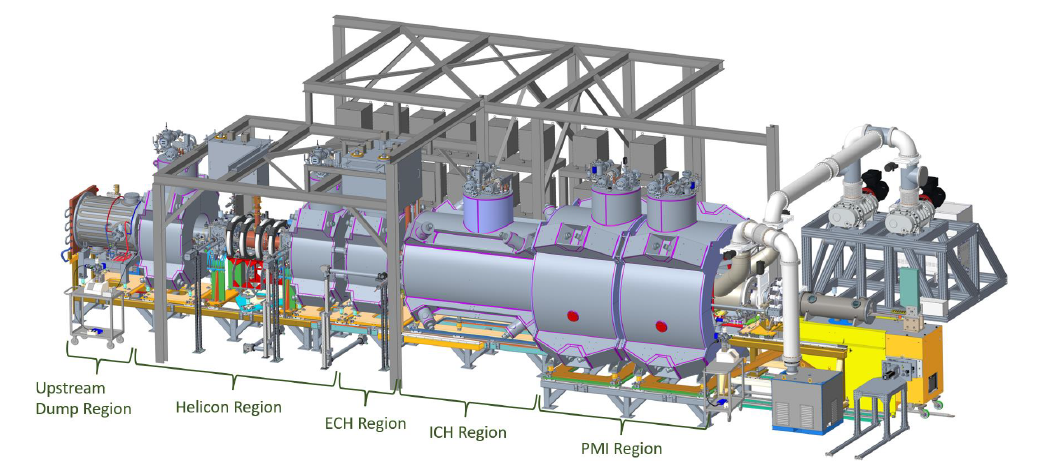}
  \caption{Drawing of the Material Plasma Exposure eXperiment (MPEX) showing the Helicon plasma source, Electron Cyclotron Heating (ECH), Ion Cyclotron Heating (ICH) and Plasma Material Interaction (PMI) regions }
\label{fig:MPEX}
\end{figure}

\mainmatter
%%%%%%%%%%%%%%%%%%%%%%%%%%%%%%%%%%%%%%%%%%%%%%%%%%%%%%%%%%%%%%%%%%%%%%%%%%%%%%%
% DOCUMENT CONTENT
%%%%%%%%%%%%%%%%%%%%%%%%%%%%%%%%%%%%%%%%%%%%%%%%%%%%%%%%%%%%%%%%%%%%%%%%%%%%%%%
\section{INTRODUCTION AND BACKGROUND}

All magnetically confined plasma fusion power plant concepts (Tokamak, Spherical Tokamak, Stellarator, Mirror, ...) must exhaust the heat and plasma from the core confinement region
to the material walls. The primary channel for this exhaust is through a plasma divertor which directs plasma along open magnetic field lines to a material target. The Material
Plasma Exposure eXperiment (MPEX) \cite{rapp2023finaldesign} illustrated in Figure 1, is a high-power, steady-state linear plasma device designed
to produce the plasma material interaction (PMI) conditions of the divertor of future magnetic confinement fusion power plants: energy flux $\mathrm{20 MW/m^{2}}$, ion fluence $\mathrm{10^{31}/m^{2}}$, pulse duration $\mathrm{10^6 sec}$. These goals of plasma exposure in MPEX are well beyond those achieved in magnetic fusion experimental devices. Successfully achieving these high power steady state
conditions for long pulses requires operational control of the heating and particle sources and the plasma flux to the walls and target. The MPEX AI Hot Spot Controller, proposed
in this project, will help achieve the operational milestones of MPEX.

The MPEX device will begin commissioning at the end of FY26. A smaller proto-MPEX was operated for 14,666 plasma discharges and will resume operation in September of 2025 as proto-MPEX-lite, with reduced capability, to test a new window for the Helicon plasma source. The proto-MPEX data has undergone surrogate modeling with machine learning methods (R. Archibald, 2022 IEEE International Conference on Big Data). This proto-MPEX data will be used to begin development of the AI digital twins described in this white paper. 

The scientific mission of MPEX is to qualify materials of different composition for use in the high energy and plasma flux conditions of a fusion power plant. The materials exposed in MPEX will in some cases be exposed to high neutron fluxes at other ORNL facilities to measure the changes to their PMI properties. The targets exposed in MPEX will be transported under vacuum to a Surface Analysis Station (SAS). The SAS will be equipped with the following diagnostics: Focused Ion Beam (FIB) for trench milling, 100-400 angstrom resolution scanning electron microscope (SEM), surface mapping x-ray spectrometer, high resolution camera, and a future upgrade to a laser induced breakdown spectroscopy quadruple mass spectrometer (LIBS-QMS). The MPEX experiments will generate diverse pre- and post-exposure measurement data of detailed material properties down to the crystal grain level in 3D for post-exposure assessment of PMI damage (e.g. cracking, melting, erosion and redeposition of the material). Physics models for the PMI, and how the material composition and manufacturing impact its performance under high energy plasma exposure, need to be validated with MPEX data to guide the selection of new candidate materials. 

Our vision for the \textit{MPEX AI Digital Twins} project is to supply experimental and physics model simulation data to train Artificial Intelligence (AI) models for data processing, analysis, operational control, PMI and materials simulation to maximize the scientific output of the MPEX device. Ultimately, an AI digital twin of MPEX material assessment metrics for tested and synthetic material types with simulated PMI will be trained by the AI Modeling Teams on the experimental and physics simulation data submitted to the American Science Cloud by this project. A purely empirical search for the best material is inefficient given the finite number of samples that can be tested on MPEX. In order to expand the material properties database for training the MPEX Material Assessment AI Digital Twin, and to gain physics understanding of the PMI processes, physics models of the material properties and PMI processes are required. The physics simulations provide detailed simulation data, like impact angles for plasma ions, sputtering yields, transport of the ionized sputtered target material in the plasma, and redeposition locations. This simulation data expands the measurement data for deeper physics understanding. The experimental data is essential to validate the PMI and material structure simulation models. The validated models can then be used to generate new simulation data of MPEX material assessments for synthetic material compositions that have not been exposed in MPEX. These predictive simulations, plus the whole experimental dataset, will be used to train the MPEX Material Assessment AI Digital Twin allowing a rapid generative AI search for new materials with reduced PMI damage by interpolating the domain of the training set. These new optimum materials can be simulated with the physics codes and/or tested in MPEX. The ability of AI neural networks to interpolate multi-dimensional parameter spaces and generate virtual data is exploited for a more efficient search for optimum materials. 

The advent of the Transformational AI Models Consortium (TAIMC) is an opportunity to engage with state of the art private and public AI developers to achieve the goals of the AI digital twins and AI accelerated physics models proposed in this project. Our partners at ORNL from the Advance Scientific Computing Research (ASCR) organization will collaborate in accelerating the integrated plasma material interaction simulation framework.   This simulation framework will provide a platform for generating simulation data across a range of physical fidelities, including hybrid methods that produce multi-fidelity results.  This data will be leveraged for AI model development, both for generation of surrogates and the automation of simulation campaigns.  A part of the research below will include collaborative efforts with the TAIMC to (i) adapt data storage approaches to ensure AI-readiness, (ii) provide a protypical exemplar to inform and exercise constructed workflows, and (iii) generate and share data, using the TAIMC unified AI data standard, for foundational models that will be trained from multiple sources across the DOE complex.  We will also collaborate with the TAIMC, as well as the planned AI modeling teams, to develop approaches for reducing the cost of data generation.  These include tailored multi-fidelity approaches as well as fine-tuning strategies to augment general, large-scale foundational models.

\clearpage
\section{MPEX DATA PROCESSING AND AI IMAGE FEATURE ANALYSIS}
\subsection{DATA PREPARATION FOR THE AMERICAN SCIENCE CLOUD}
\textbf{Approach}: Both experimental measurement data and physics model simulation data will be organized and processed for transfer to the American Science Cloud. Standard data schema will be defined for MPEX and used to organize both experimental and simulation data in collaboration with the TAIMC. The data obtained from the proto-MPEX device (https://mpex.ornl.gov/proto-mpex/) is already on the data server that will be used by MPEX. Starting in September of 2025 a new run of data from proto-MPEX with somewhat reduced capability (proto-MPEX-lite) will begin operation and produce new data. The proto-MPEX data will be the first cohort to be processed and uploaded to the American Science Cloud. The physics simulation and analysis codes will be integrated with the ORNL developed IPS framework \cite{Elwasif:2010f}. The simulation codes are wrapped in IPS modules and run in containers on high performance computing resources. Input and output to the codes is performed by standardized statefiles similar to the IMAS data schema used for the ITER project \cite{Imbeaux_2015}. The IPS framework allows coupled code workflows to be built and executed. The IPS statefile structure is flexible and can be adapted to the requirements of the American Science Cloud. For the experimental data, we will deploy a data backbone that links raw images, metadata, AI outputs, and twin states so results flow seamlessly from the experiment to analysis and physics simulation validation and uncertainty quantification. Images and manifests will be ingested into versioned object storage, embedded calibration, and indexed metadata for sample provenance and exposure conditions. AI inferences (masks, features, uncertainties) will be stored in columnar formats (HDF5) and referenced to each image; calibrated twin parameters and summary stress/temperature fields will be stored as versioned artifacts. This will constitute the MPEX curated experimental database, which will be released as an open-source database through The American Science Cloud. This database can be leveraged for choosing future target and exposure conditions for future experiments and complimentary to lifetime assessment of materials in other plasma and high heat flux facilities. Overall, this MPEX data curation will convert images and logs produced by experiments into operational intelligence: standardized metrics for PMI damage assessment, and quantitative lifetime forecasts. The MPEX Materials Assessment AI Digital Twin, described in Section 6, trained on this database, will enable retrospective meta‑analysis across materials and geometries, accelerating materials down‑selection and informing design of next‑generation targets and fixtures. 

\subsection{DEVELOP AI-DRIVEN EVALUATION FRAMEWORK FOR TARGET DATA}

\textbf{Approach}:  This project will develop a data-driven pipeline that ingests pre‑ and post- plasma exposure high resolution images and outputs quantitative morphologies with calibrated uncertainties. An ongoing LDRD image‑processing effort (PI: Juneja) has established a robust pipeline for microstructure understanding in electron-beam high heat flux exposed materials. In collaboration with the TAIMC experts, we will adapt and extend the AI models, data engineering, and image processing methods to MPEX target images and other SAS measurements. The focus of image analysis will be on crack detection and segmentation (length, width, branching, density, orientation, fractal dimension), surface melting and resolidified features (bead geometry, area fraction, depth proxies), and redeposition and roughness surrogates (areal coverage, texture metrics, thickness proxies). Starting with a small, labeled seed set, we will use self‑supervised pretraining to capture common texture/contrast cues in MPEX data, then fine‑tuned models with active learning so the algorithm flags high‑uncertainty regions. The inference stack will output morphology masks and derived metrology - crack length/width/orientation and density; melt area fractions and bead geometry; redeposition coverage and texture metrics - together with calibrated uncertainty, passing any low‑confidence cases to an operator for adjudication. Quality assurance will be embedded through automated checks for scale‑bar consistency, focus/illumination, and glare, and through synthetic robustness tests (blur/contrast/noise). The result will be an operator‑in‑the‑loop, evaluation tool that produces standardized, quantitative features from every target SAS data set.

\subsection{CUSTOMIZE DATA INTERFACE USING THE NOVA-GALAXY FRAMEWORK}

\textbf{Approach}:  NOVA (https://github.com/nova-model) is a web-based python framework to enable the analysis and visualization of scientific data for materials scientists developed by the Application and Software Engineering groups at ORNL. NOVA leverages the Galaxy Project (https://galaxyproject.org/), a platform for creating scientific workflows for data analysis that provides a suite of customizable tools that can be assembled into workflows to access data, run analysis codes on remote HPC computers, fetch output results, and perform basic visualization tasks. When combined together NOVA-GALAXY vastly simplifies the development and deployment of sophisticated user interfaces that employ advanced visualization techniques, while at the same time leveraging the workflow and data management capabilities needed to create cutting-edge AI-enabled HPC applications. The MPEX data and analysis interface will be configured to connect to the American Science Cloud.

\section{MPEX PLASMA MATERIAL INTERACTION SIMULATION FRAMEWORK}

The MPEX PMI physics simulation framework will integrate advanced kinetic plasma simulations with multiscale PMI modeling to provide a predictive, experimentally validated, framework for simulating MPEX operations with emphasis on the divertor target. Because PMI processes span temporal and spatial scales—from atomistic sputtering to device size impurity transport, a unified environment is required that couples plasma transport, RF heating effects, impurity generation, and material response across multiple physics fidelities. To meet this need, ORNL has developed, as a part of internally funded LDRD and RF-SciDAC projects, the Simulated Transport of RF Impurity Production and Emission (STRIPE) framework \cite{kumar:2025}, which integrates plasma transport, RF heating and sheath modeling, ion–material interactions, and impurity transport. STRIPE has already been applied to linear devices (Proto-MPEX \cite{Rapp:2024}, PISCES-RF \cite{Dhamale:2024}) and tokamaks (WEST \cite{kumar:2025,Tierens:2024}, DIII-D), and has been used to predict RF impurity sources for ITER. 

\begin{figure}
    \centering
    \includegraphics[width=\textwidth]{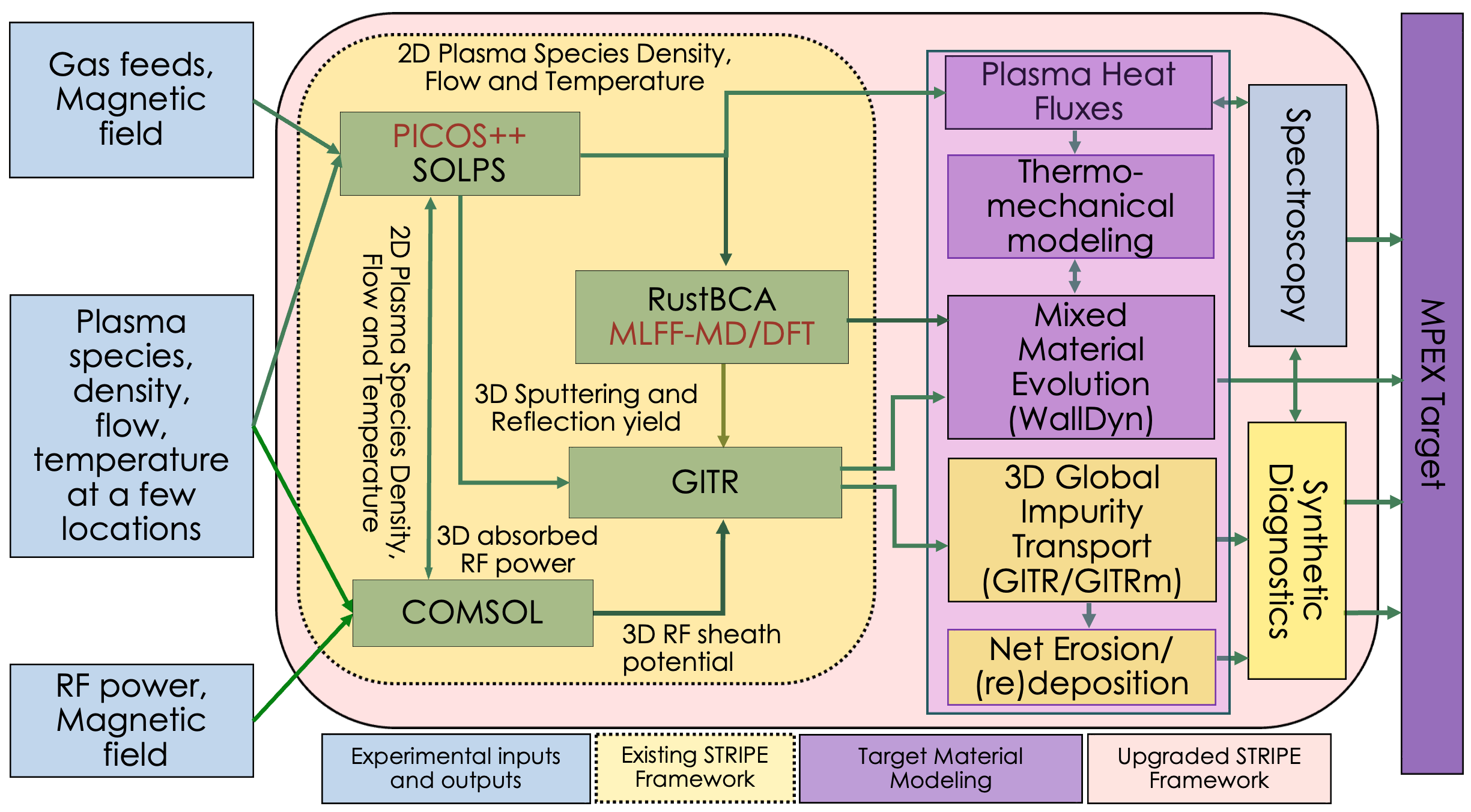}
    \caption{Enhanced STRIPE framework for MPEX modeling, incorporating the upgraded PICOS++ kinetic solver, COMSOL RF sheath modeling, advanced sputtering calculations (BCA, MLFF-MD, DFT), and GITR/GITRm Monte Carlo impurity transport simulations, integrated with synthetic diagnostics for predictive divertor-target PMI analysis.}
    \label{fig:stripe_framework}
\end{figure}

To make STRIPE a truly predictive tool for PMI studies at the MPEX divertor, the workflow (Fig.~\ref{fig:stripe_framework}) will incorporate an upgraded 2D kinetic particle in cell solver (PICOS++), advanced sputtering models (BCA, MLFF-MD, DFT), COMSOL-based RF heating and sheath calculations \cite{Beers:2021, Beers2, Tierens:2024}, and AI-accelerated GITR/GITRm transport \cite{Younkin:2021, Nath:2025, nath:2023}. Together, these capabilities will enable quantitative modeling of erosion, impurity sourcing, and redeposition under reactor-relevant heat and particle fluxes at the MPEX divertor target.

\subsection{AI ACCELERATED KINETIC PLASMA SIMULATION FOR MPEX}

\textbf{Approach}: PICOS++ (Particle-In-Cell for Open Systems) \cite{Kumar:2022, kumar:2023, Kumar:2024, Islam:2024} will serve as the primary kinetic engine for modeling MPEX’s multi-frequency RF heating environment. Developed at ORNL and validated in Proto-MPEX and PISCES-RF, PICOS++ includes MPI+OpenMP parallelization, RF-heating modules, Fokker–Planck collision operators, volumetric particle sources, and mirror trapping physics. These capabilities will be extended to capture the combined effects of helicon, EBW–ECH, and ICH in MPEX’s magnetic mirror geometry, with explicit attention to how RF-driven non-Maxwellian particle distributions enhance ion and electron fluxes to the divertor.

Upgrades will include velocity-dependent forcing functions and spatially resolved RF power deposition profiles—derived from GENRAY-C and AORSA for ECH \cite{Caneses:2021} and COMSOL-based full wave solvers for helicon \cite{Islam:2023} and ICRH \cite{Goulding:2023}—enabling accurate simulation of anisotropic heating and the fast ion/electron populations responsible for enhanced divertor-target erosion.

While the ultimate objective is a fully 2D (radial $\times$ parallel) kinetic solver, PICOS++ currently operates in a flux-surface-resolved mode, evolving distribution functions along field lines and deriving transport coefficients such as diffusivities, convective velocities, and confinement timescales. Coupled with machine learning/AI methods and  experimental datasets, this approach will allow reconstruction of 2D plasma profiles from flux-tube simulations. The resulting profiles will directly inform impurity sourcing and redeposition models at the MPEX divertor target.

To ensure computational tractability, several acceleration strategies will be employed:
\begin{itemize}
    \item {\textbf{Time-splitting} to sub-cycle electron dynamics while updating field solutions on longer time steps.}
    \item {\textbf{GPU acceleration} using the ORNL's domain specific compiler Graph Framework \cite{cianciosa:2025} for particle orbit calculations and for ECH ray tracing.}
    \item {\textbf{Normalizing flows} can significantly improve velocity-space sampling efficiency and reduce statistical noise in high-energy tail populations by learning and transforming from simple latent distributions (like standard Gaussians) to complex velocity distributions that naturally capture rare high-energy events. The Pseudo-Reversible Normalizing Flow (PR-NF) \cite{Yang:2024, yang2025conditional} is well-suited for the velocity configuration space because it uses fully-connected neural networks to map latent variables to velocity space without imposing hard architectural constraints that limit network expressivity, unlike traditional invertible architectures. Once trained, the model can efficiently generate velocity samples by drawing from the simple base distribution and transforming them through the learned mapping, effectively concentrating computational effort on producing samples in regions of interest rather than rejecting the majority of samples as in traditional acceptance-rejection methods. This approach reduces the variance in Monte Carlo estimates of high-energy populations by orders of magnitude, as it generates samples proportional to the target distribution rather than relying on rare acceptance events, while the neural network's ability to capture complex correlations enables accurate representation of non-Maxwellian features and anisotropic distributions commonly encountered in plasma physics applications.}
\end{itemize}

This combination of physics fidelity, numerical acceleration, and reduced-order modeling will provide MPEX with a predictive, high-performance kinetic simulation capability. Coupled with realistic RF power deposition from ray tracing and full-wave solvers, the enhanced PICOS++ will quantify the role of non-thermal particles in impurity sputtering, sheath potentials, and PMI under reactor-relevant conditions.

%{\color{red}{Need to add something about PICOS++ needing neutral profiles, but kinetic codes e.g., EIRENE are too computationally expensive.}}
\subsection{KINETIC FLUID HYBRID PLASMA AND NEUTRALS MODEL}

\textbf{Approach}: The MPEX plasma is highly collisional but also has non-maxwellian features like anisotropic pressure parallel and perpendicular to the magnetic field due to magnetic mirror forces and ECH and ICH heating. The PICOS++ model is able to simulate the kinetic plasma features but it does not have neutral particle species.  Adding kinetic neutrals to PICOS++ would increase the computational burden prohibitively. The C1 code solves a set of Braginskii-like fluid moment transport equations in 1D, for ion and neutral densities, ion and electron temperatures, and the parallel ion velocity. C1 uses an MHD equilibrium solution to automatically generate a grid that conforms to the magnetic configuration, and is currently being developed to include simplified radial particle and energy fluxes. The plasma transport in MPEX is predominantly along the magnetic field so a 1D model is a good start. The C1 code is a 1D reduction of the 2D finite volume code C2 that already includes the flux surface normal direction. A kinetic-fluid hybrid model will be developed that couples PICOS++ and C1 treating C1 as the equations for the plasma and neutral density, temperature and flow velocity moments and PICOS++ computing the higher velocity moment departures from a flowing Maxwellian (Macro-Micro Decomposition). Coupling of the particle in cell code PICOS++ and the fluid code C1 with fluid neutral will be demonstrated in 1D and then generalized to 2D in space.

\subsection{ADVANCED STRIPE FRAMEWORK FOR PMI MODELING}

\textbf{Approach}: The STRIPE framework will provide the multiscale linkage from plasma heating conditions to divertor-target erosion/re-deposition and impurity transport. In MPEX STRIPE will model simultaneous ion- and electron-driven erosion, sputtering, and redeposition. In addition to direct divertor erosion, impurity generation from other components—such as helicon antenna windows—can significantly affect plasma conditions at the divertor, contributing to enhanced erosion, redeposition, and target-surface evolution. Proto-MPEX experiments have already demonstrated that RF sheath rectification accelerates ions to keV energies, causing severe erosion of AlN windows \cite{Beers:2021,Beers2}. In MPEX, these mechanisms will be equally critical at the divertor plate, where sputtered atoms are promptly ionized and transported, degrading plasma purity and limiting material lifetime.

MPEX operates in an open, non-uniform magnetic configuration with sequential, multi-frequency RF heating stages: helicon excitation at 13.56~MHz (up to 200~kW) \cite{LUMSDAINE2020112001, Caneses:2021}, EBW mode–converted electron cyclotron heating (ECH) at 70~GHz (up to 400~kW) \cite{Marin_2022}, and ion cyclotron heating (ICRH) at 4–9~MHz (up to 400~kW) \cite{Beers2018, Goulding:2023}. This heating combination produces high-density deuterium plasmas with divertor-target heat fluxes approaching $20$~MW/m$^2$ and pulse durations of up to $10^6$~s, directly exposing candidate materials to reactor-relevant operating conditions. 
% Figure~\ref{fig:edgeheating} highlights strong edge power deposition in MPEX, a driver of impurity generation and transport toward the divertor.

MPEX is specifically designed to test plasma-facing materials beyond tungsten, including advanced ceramics, composites, and alloys as well as neutron irradiated materials. To model this broad materials space, STRIPE will be extended with both physical and chemical sputtering models, leveraging molecular dynamics (MD) and density function theory (DFT) calculations to capture temperature- and flux-dependent erosion mechanisms. MD with machine-learned force fields (MLFFs) trained on DFT data will enable predictive sputtering yields for complex ceramics and alloys, validated against dedicated Proto-MPEX, PISCES-RF, and future MPEX experiments. Unlike current empirical approaches such as RustBCA, which are limited to traditional material surfaces due to scarce experimental datasets, MLFF–MD/DFT-based models can generalize to new and non-traditional materials, providing a path to predictive sputtering simulations for future reactor applications. For mixed-material targets, STRIPE will integrate the WallDYN code \cite{Schmid:2018} to capture evolving surface composition and multi-species impurity transport under both steady-state and transient operating conditions.

Through this integration of atomistic models, kinetic plasma simulations, and impurity transport, STRIPE will deliver a predictive, AI-accelerated PMI modeling capability directly tied to MPEX’s divertor target program. This unified framework will connect plasma heating scenarios, impurity sourcing from multiple components, and material degradation in a simulation–experiment feedback loop, supporting experimental planning in MPEX campaigns. 

\subsection{TURBULENCE MODELING OF MPEX}
\textbf{Approach}: LLNL will perform high-performance 3D simulations of turbulence in the MPEX device. These will provide time series of fluctuating heat and particle loads to material surfaces, and quantify the impact of plasma fluctuations on plasma-neutral interactions. We will use a multi-fidelity approach using the kinetic code COGENT and fluid code BOUT++/Hermes-3.

COGENT: The gyrokinetic continuum code COGENT has been effectively applied to linear plasma devices, such as magnetic mirror configurations, yielding significant computational speedups while retaining high-fidelity physics modeling. COGENT was recently extended with improved sheath boundary conditions to model plasma-wall interactions \cite{Geyko:2024}, further enhancing its relevance to MPEX where sheath physics strongly influences particle and heat fluxes to the target plates. The COGENT kinetic model will capture trapped particle effects, non-Maxwellian ion distributions, and kinetic instabilities.

BOUT++/Hermes-3 is a high-performance drift-reduced plasma fluid turbulence model \cite{Dudson:2024} that has been applied to simulations of linear devices, including studies of the interaction of turbulent plasma with neutral gas \cite{Leddy:2017} showing that turbulence enhances the effective atomic rates. Hermes-3 describes drift-wave and rotation-driven interchange turbulence, coupled to a fluid neutral model for self-consistent description of the particle and heat sources and sinks.

Extension: As a longer-term objective, we may also consider fully kinetic simulations involving both ion and electron dynamics in COGENT, although these would be significantly more computationally intensive. Such studies, however, could serve as valuable benchmarks for both ML-driven ROMs and reduced-fluid models under conditions of strong non-Maxwellian features, making them critical for cross-validation in advanced modeling workflows.

Additional ML component: To address the high-dimensional and nonlinear nature of plasma turbulence in MPEX, we are exploring the integration of machine learning (ML) and artificial intelligence (AI) tools to develop reduced-order models (ROMs). In particular, we are investigating latent space dynamics identification (LaSDI) frameworks \cite{Bonneville:2024}, which learn low-dimensional manifolds where plasma dynamics can be more efficiently simulated. These models aim to capture essential features of phenomena such as drift-wave turbulence and intermittent transport without resolving the full kinetic detail at every time step, enabling rapid predictive modeling across a range of operational regimes. Beyond turbulence modeling, ML can also enhance multi-fidelity simulations by guiding the placement of expensive kinetic COGENT runs to calibrate and validate UEDGE fluid simulations, particularly in regimes where fluid assumptions break down (e.g., strong anisotropy or kinetic sheath effects). This hybrid approach could lead to more robust UEDGE surrogate models, improving extrapolation capability and uncertainty quantification in edge-plasma simulations for MPEX.

\section{MPEX AI HOT SPOT CONTROLLER}

\textbf{MPEX control requirements}

The plasma flux to the walls in MPEX needs to be controlled to avoid hot spots that can crack the Helicon and ICH antenna windows. The windows are monitored by real-time cameras during operation. The coil currents in MPEX can be adjusted to direct the flow of heated plasma towards the target chamber and away from these windows. The position of the magnetic field where resonant absorption of the ECH and ICH power occurs can be adjusted by changing the coil currents but the magnetic fields also impact the plasma particle orbits that can impact the walls or windows. The rectified sheath in front of the Helicon antenna produces high voltage radio frequency oscillations and asymmetric PMI with the antenna window \cite{Beers:2021}. The MPEX magnet system consists of twenty-three coils in total, with nineteen superconducting coils and four resistive copper coils \cite{burkhardt2023mpexmagnets,rapp2023finaldesign}. Determining the optimum coil currents by trial and error is not efficient for control.  We propose to train an AI model that will estimate the plasma heating of the walls for a given magnetic field and heating source configuration. This AI model can inform the operations control what combination of coil currents will direct the plasma to the target with minimal hot spots due to plasma flux on the windows and walls. Since proto-MPEX is now ready for operations, this AI Hot-Spot controller will first be trained using proto-MPEX data.

\subsection{AI HOT-SPOT IDENTIFICATION OF CAMERA REAL-TIME IMAGES FOR A RANGE OF PLASMA CURRENTS AND HEATING}
\textbf{Approach}:
MPEX will have 7 fast, low resolution, cameras pointed at various solid-material objects and plasma along the length of the device. One view is of the MPEX target.  The cameras can frame at up to 18k fps.  These are full-frame cameras, i.e. not-interleaved imaging, so that each frame can be clearly exposed and time correlated, in order to synchronize with other timing signals, such as a triggering event.  The cameras have a rolling-buffer of memory so when an event happens it has recorded images both after and BEFORE the triggering event. We propose to use AI image processing to identify hot spots in the camera images. The magnet coil currents will be varied in MPEX at low power operation to produce a database of hot spots for each coil current settings. Real time observation of a dangerous hot spot will trigger a warning. 

\subsection{SIMULATE HOT SPOTS ON MPEX WALL WITH PMI SIMULATION FRAMEWORK}
\textbf{Approach}: The hot spot locations will be simulated with the plasma and heating models of the STRIPE framework. Synthetic diagnostic images at the camera location will be post-processed from the simulations. The location of hots spots in the simulations on the walls of MPEX, outside of the camera views, will also be identified. Variations in the coil currents at both low and high heating power will be simulated to provide hot-spot data that overlaps the measured camera data at low power and extends the simulated data to high power and regions not observed by cameras.

\subsection{TRAIN AN MPEX AI HOT SPOT CONTROLLER}
\textbf{Approach}: In collaboration with the AI Modeling Teams, AI models will be trained to the hot spot database, stored in the American Science Cloud, using the coil currents and heating powers as inputs and the locations and intensity of hots spots as output. The AI does not need to know any physics. This MPEX AI Hot Spot Controller is used in operations by using the AI image processed camera data to warn when hot spots are observed. If the operator determines that there is a dangerous hot spot on the antenna windows they can ask the AI Hot Spot Controller to find a set of coil currents that will reduce the hot spot, without producing new ones elsewhere on the walls, and maximizing the energy flux to the target (cost function minimization). The operator can then adjust the complex set of coil currents and heating sources as advised by the AI model. The location of the magnetic field resonance 2nd harmonic absorption for the ECH should be constrained to be a fixed point in the coil adjustments. The MPEX AI Hot Spot Controller will be integrated into a real time MPEX operations control system to help achieve the extremely long pulse high power exposure goal. The primary metrics of impact by the AI Hot Spot Controller is if the MPEX operators can reach high power and long pulse goals faster than they would have without the AI Controller.

\section{MPEX TARGET MATERIALS SIMULATION}
\subsection{VALIDATE THERMOMECHANICS SIMULATIONS OF MPEX TARGET MATERIAL PROPERTIES}
\textbf{Approach}: The MPEX SAS data will be employed to validate the materials response under MPEX operating regimes. The SEM microstructure data will be an input to multi-scale physics models to predict the thermomechanical response of materials, including crack initiation and propagation, erosion, and thermal performance. %Validation of the meso-scale models will be conducted through comparisons with MPEX SEM microstructural data. 
By matching the observed damage observables such as crack densities, orientations, and sub-grain-scale deformation patterns, these models will be refined to ensure close alignment with experiments. In conjunction, continuum simulations will also be leveraged to understand macro-scale effects, including material response at the meso-scale as input constitutive behavior. Commercial finite‑element software (ANSYS/COMSOL) will be used to simulate transient heat flow and elastic‑plastic mechanical response under recorded MPEX exposure loads and cooling boundary conditions. 
The ORNL-developed CabanaPD fracture mechanics software will offer an additional option not only to predict thermomechanical response, but also to capture thermomechanically driven failure~\cite{cabanapd}. CabanaPD employs a peridynamics framework that directly simulates crack initiation and propagation. Furthermore, CabanaPD computations use a meshfree approach, which overcomes the challenges of mesh-based methods associated with constraining crack propagation to the underlying mesh, making CabanaPD highly effective in capturing advanced fracture phenomena, such as fragmentation and complex crack patterns. Although it is a more computationally expensive method, CabanaPD runs on both CPUs and GPUs across diverse hardware architectures, enabling scalable, high-resolution simulations. CabanaPD has been shown to capture fracture due to thermomechanical shock loading. We will use CabanaPD to compute key macro-scale fracture outputs that result from high-pulsed plasma loads and thermal cycling under MPEX-relevant conditions.

% Continuum validation??

AI‑identified PMI damage features will be mapped to physically meaningful state variables (e.g., principal stress maxima, plastic strain accumulation, temperature‑time integrals, and as appropriate degree of fracture) using surrogate regressors and informed priors and then used in a Bayesian assimilation loop (ensemble variational methods) to calibrate uncertain material and interface parameters. This closes the observation‑model gap and yields posterior distributions for damage‑relevant quantities, enabling estimation of remaining useful life and scenario forecasting for proposed future exposures. We will explicitly account for property shifts from redeposition and roughness evolution by updating effective surface/emissivity, and we will validate consistency by checking predicted crack spacing and melt footprints against measured morphologies on held‑out targets.

\begin{figure}[h]
    \centering
    \includegraphics[width=\textwidth]{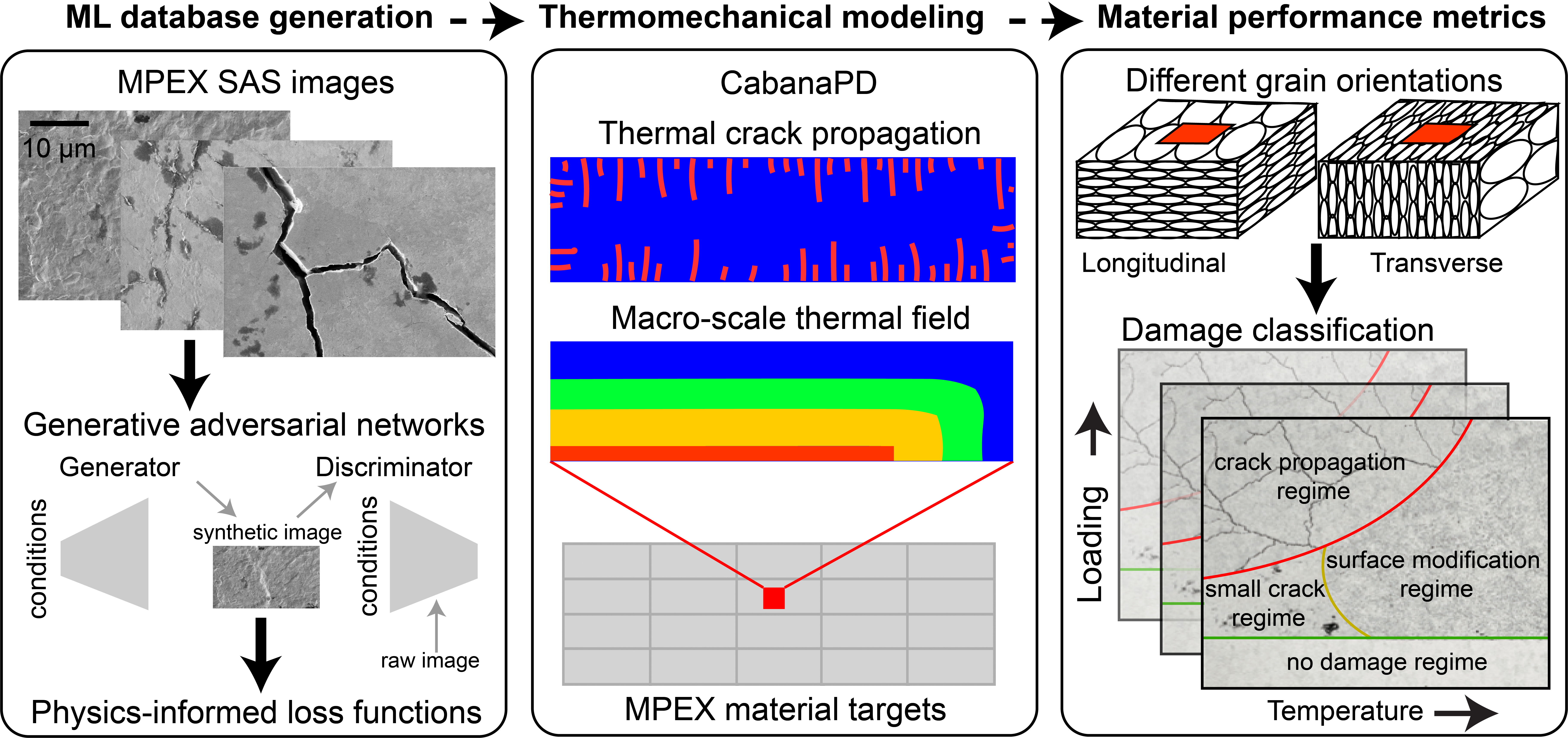}
    \caption{Materials Simulation workflow for high heat flux electron beam facility showing machine Learning (ML) image database generation (left) thermomechanical modeling with CabanaPD (middle) and material performance assessment metrics (right)}
    \label{fig:MPEX-PMI}
\end{figure}

A workflow incorporating the AI image processing, thermomechanical modeling and material performance assessment metrics has been developed for high heat flux testing of tungsten in electron beam facilities as an ORNL funded (LDRD) project. This workflow, illustrated in Fig.~\ref{fig:MPEX-PMI}, is a prototype of the one proposed for MPEX. 

\subsection{DEFINE MATERIALS METRICS FOR PMI ASSESSMENT}
\textbf{Approach}: We will define material state performance metrics that will capture the pre-exposure state of each target and serve as the input to the AI Material Assessment Digital Twin. The performance metric will comprise statistically robust descriptors of grain size, distribution, grain-boundary texture, porosity fractions, surface defect density, and tensor properties such as elastic moduli and hardness. The Physics-constrained high-fidelity thermo-mechanical models developed in section 5.1 will generate training data and inform fast surrogates that map material state performance metrics along with exposure conditions such as incident flux, temperature, duration conditions to a damage state metric quantifying PMI outcome (e.g., erosion depth, crack density, redeposition thickness, roughness, and property degradation). Rigorous uncertainty quantification, verification and validation, and robustness testing will provide calibrated confidence intervals and define operating envelopes by bounding exposure conditions that keep material components below specified thresholds guided by these material performance metrics. The resulting models will enable comparative, real-time performance forecasting across designs and will be disseminated through the MPEX database.

\section{MPEX MATERIAL ASSESSMENT AI DIGITAL TWIN}
\textbf{Achieving the mission objective of MPEX}: Finding and testing better materials for fusion power plant divertors can be accelerated by training an MPEX AI Material Assessment Digital Twin that takes metrics of material properties and plasma exposure meta-data as input and outputs the PMI damage assessment in reduced metrics like physical and chemical erosion rates, melting threshold, redeposition and co-deposition rates, cracking characteristics and thresholds, etc.. This is an AI Digital Twin of the final MPEX PMI damage assessment metrics from both experiments and the physics models of the PMI and materials. Training this MPEX AI Digital Twin, on both experimental and synthetic physics simulation data, enables the AI to search for optimum material candidates for a fusion power plant divertor using the multi-dimensional interpolation power of neural networks. 

The primary focus of this project is to generate and validate the training data. However, we will collaborate heavily with the TAIMC and the AI model teams in the development of the Digital Twin.  This development will leverage AI-based surrogates trained by data from the physics-simulations discussed above. It will also leverage AI-driven design strategies to efficiently sample the high-dimensional search space using the simulation framework to drive experiments and the experimental data back to the simulation models to correct for bias.  In addition, the data generated during a Digital Twin Campaign will be processed for AI-readiness and made available to a more general science community via the American Science Cloud.

\subsection{VALIDATE THE PMI AND MATERIAL PHYSICS MODELS WITH CURATED MPEX DATA}

\textbf{Approach}: The experimental measurements of the damage due to PMI will be used to validate the physics models for the PMI and material properties.  Synthetic data (images of the target, erosion and redeposition rates and locations, cracking, etc..) that can be directly compared with MPEX measurements will need to be produced from the physics model simulation results. The uncertainty in the simulations needs to be quantified before using the physics models to simulate synthetic materials. 

\subsection{GENERATE SYNTHETIC MATERIAL SIMULATION TRAINING DATA}
\textbf{Approach}: Variations of the material properties well outside the experimental dataset is used to produce new simulation data for training the AI Digital Twin. These cases are not primarily chosen to yield better PMI results but to expand the training set to enable the AI Digital Twin to have a larger domain of training. Some of these synthetic materials may not be possible to fabricate but that is not a constraint for this purpose. 

\subsection{TRAIN THE MPEX MATERIAL ASSESSMENT AI DIGITAL TWIN ON THE COMBINED EXPERIMENTAL AND SIMULATED DATA}

\textbf{Approach}: In collaboration with the AI Modeling Teams, an AI Digital Twin of the input materials property metrics and output PMI damage assessment metrics, stored in the American Science Cloud, will be built and trained. After training the Digital Twin on the combined experimental and simulated data sets, improved PMI performance candidates will be found by the Digital Twin interpolating the multi-dimensional data. These candidates can then be simulated with the physics codes or, if they can be easily manufactured, tested in MPEX. This is an iterative process of refining the Digital Twin and looking for the best candidate materials. We propose to start with a simpler dataset of tungsten damage from an electron beam heating experiment that has been collected as part of an ORNL LDRD project (Rinkle Junja PI). This dataset focuses on cracking patterns due to thermal stress for tungsten targets with different grain structures. A prot-type AI Digital Twin of this much lower dimensional dataset would be a good first test of the required AI model designs and how they would scale up to the higher dimensional proto-MPEX and MPEX data. Physics model simulation of the electron beam system will add to the training set. ORNL has plans to build an electron beam facility in the MPEX building and Type One Energy plans to build one at their Bull Run site in Oak Ridge. These facilities will benefit from the physics simulation capability and AI Digital Twin models of the existing electron beam data. The primary metric of impact for the MPEX AI Material Assessment Digital Twin is if the AI can find new candidate materials to test in MPEX that would not have been identified by the research team. 

% \nocite{*}
%\bibliography{mpex}% Produces the bibliography via BibTeX.
%\bibliographystyle{unsrtnat} % or plainnat with numbers
\printbibliography

\end{document}